\documentclass{elsart}
\usepackage{amssymb}
\usepackage{wasysym}

\begin{document}
\input psfig.sty
 
\hoffset=0pt
\voffset=0pt
\hyphenation{cal-o-rim-e-ter}
\hyphenation{ped-es-tal}
\hyphenation{gain--mon-i-tor-ing}
\hyphenation{mon-i-tor-ing}

\begin{frontmatter}
 
\title{\bf The automatic gain-matching \\
in the PIBETA CsI calorimeter}
 
\author[UVa]{E.~Frle\v{z}\thanksref{author}},
\thanks[author]{Corresponding author. Tel: +1--434--924--6786;
fax: +1--434--924--4576. \\ {\sl E--mail address:} 
frlez@virginia.edu (E.~Frle\v{z}).\hfill}
\author[UVa]{M.~Bychkov}, and
\author[UVa]{D.~Po\v cani\'c}
\address[UVa]{Department of Physics, University of Virginia,
Charlottesville, VA~22904, USA}
 
\begin{abstract}                       
Segmented electromagnetic calorimeters are used to determine both the
total energy and direction (momentum components) of charged particles
and photons.  A trade off is involved in selecting the degree of
segmentation of the calorimeter as the spatial and energy resolutions
are affected differently.  Increased number of individual detectors
reduces accidental particle pile-up per detector but introduces
complications related to ADC pedestals and pedestal variations,
exacerbates the effects of electronic noise and ground loops, and
requires summing and discrimination of multiple analog signals.
Moreover, electromagnetic showers initiated by individual ionizing
particles spread over several detectors. This complicates the precise
gain-matching of the detector elements which requires an iterative
procedure.  The PIBETA calorimeter is a 240-module pure CsI
non-magnetic detector optimized for detection of photons and electrons
in the energy range 5--100\,MeV.  We present the computer-controlled,
automatic, in situ gain-matching procedure that we developed and used
routinely in several rare pion and muon decay experiments with the
PIBETA detector.  \par\noindent\hbox{\ }\par\noindent {PACS Numbers:
29.40.Gx, 29.40.Vj, 29.85.+c} \par\noindent\hbox{\ }\par\noindent {\sl
Keywords:}\/ Large acceptance segmented calorimeter; Pure CsI
scintillators; Detector calibration gain-matching and performance
\vfill
\end{abstract}
\end{frontmatter}
\vfill\eject
 
\section{Introduction}\label{sec:int}
\medskip              
\renewcommand{\theenumi}{(\arabic{enumi})}
\renewcommand{\labelenumi}{\theenumi}

The PIBETA detector~\cite{Frl04a} was built at the Paul Scherrer
Institute~\cite{Psi94} for the precise measurement of the branching
ratio of the pion beta decay $\pi^+\rightarrow\pi^0e^+\nu_e$ as well
as several other rare pion and muon decays~\cite{Poc91}.  During the
production runs in 1999-2001 and 2004, we collected a large data
sample of pion and muon decay events corresponding to a total of
$2.37\times 10^{13}$ pion stops, with stopping rates ranging between
$5\times 10^4\,\pi^+/$s and $1\times
10^6\,\pi^+/$s~\cite{Poc04,Frl04b,Bych08}.

In this paper we describe the computer-controlled gain-matching
procedure used for the PIBETA calorimeter modules. The procedure was
an essential component of an almost complete experiment automation,
enabling our detector to run virtually free of human intervention for
long periods of time.

Several systems have been used over the years for calibration and gain
monitoring of detectors based on photomultiplier tubes (PMT)~\cite{Ber89,Bia91,Mit93,Zab94,Heh95,Pei96,Ada02,Han02,Tan03,Bat04,Bat06}.
A majority of experiments use artificial light sources for a fixed
energy reference, such as lasers or LED pulses. The reference signals
are usually interfaced to the individual detectors via optical
fibers. An alternative approach uses laser light to excite a single
Plexiglas plate integrated within a calorimeter box, obviating the
need for individual fiber couplings~\cite{Bra93,Frl99,Jon07}.

We have chosen to rely on the comparison between measured and
simulated energy spectra of weak pion and muon decays in order to
provide fixed energy references for the gain factors and applied high
voltage settings of individual detectors.  We have used both
continuous energy spectra having a well defined end-point, such as the
Michel decay of the muon, $\mu^+\to e^+\nu\bar{\nu}$, with maximum
positron energy of $m_{\mu^\pm}/2=52.8\,$MeV, as well as the
mono-energetic 69.8\,MeV positron peak of the rare $\pi^+\to e^+\nu$
decay, also designated as $\pi_{e2}$ decay.  These energy scales,
utilized in automatic gain-matches, were cross-checked offline for
consistency with the 135.0\,MeV energy peak position of reconstructed
neutral pions from the $\pi_\beta$ decay.

This article contains the following discussions: in
Section~\ref{sec:det} we briefly describe the general design of the
PIBETA detector. The pure CsI calorimeter, the most important part of
the detector, is described in more detail in Sec.~\ref{sec:calo}.
Section~\ref{sec:pmt} deals with the calorimeter PMT's and associated
voltage dividers, custom-made for this experiment to insure the
required linearity of response and the large dynamical range.  The
main subject of this article, the flexible gain-matching procedure, is
covered in Sec.~\ref{sec:match}.  Several gain-matching examples from
the PIBETA production runs are presented in its subsections.  The
temperature dependence of the calorimeter analog to digital converter
(ADC) spectra is illustrated in Sec.~\ref{sec:temp}.  The resulting
calorimeter energy resolution and the PMT gain stability are discussed
in Sec.~\ref{sec:resol}.  We close with the summary of the main points
and advantages of our gain-matching procedure.

\bigskip
\section{PIBETA detector}\label{sec:det}
\medskip  

The {\sc PIBETA} apparatus is a large solid angle non-magnetic
detector optimized for measurements of photons and electrons in the
energy range of 5--100$\,$MeV. The main sensitive components of the
apparatus, shown and labeled in Fig.~\ref{fig:det}, are:
\begin{enumerate}
\item a thin forward beam counter BC, two cylindrical active
collimators AC$_1$ and AC$_2$, and an active degrader AD, all made of
plastic scintillators, used for the beam definition;
\item a segmented active plastic scintillator target AT, used to stop 
the beam particles and sample lateral beam profiles;
\item two concentric low-mass cylindrical multi-wire proportional
chambers MW\-PC$_1$ and MW\-PC$_2$ for charged particle tracking,
surrounding the active target;
\item a segmented fast plastic scintillator hodoscope PV, surrounding
the MW\-PC's, used for particle identification;
\item a segmented fast CsI shower calorimeter, surrounding
the target region and tracking detectors in a near-spherical geometry;
\item cosmic muon plastic scintillator veto counters CV around 
the entire apparatus (not shown).
\end{enumerate}

All above detector components, together with the delay cables for
analog signals from photomultiplier tubes, high voltage (HV) power
supplies, MWPC instrumentation and gas system, fast trigger
electronics, two front end computers for data acquisition and slow
control system are mounted on a single platform and can be moved as a
single unit into the experimental area.  After the detector platform
is precisely positioned with respect to the beam line, providing the
electrical power and Ethernet connections makes the detector
operational in less than 24 hours.

\bigskip
\section{PIBETA calorimeter}\label{sec:calo}
\medskip

The central part of the {\sc PIBETA} detector is the electromagnetic
shower calorimeter, shown in Fig.~\ref{fig:ball}.  The calorimeter's
nearly spherical geometry is obtained by the ten-frequency class II
geodesic triangulation of an icosahedron~\cite{Ken76}.  Its active
volume is made of pure CsI~\cite{Kob87,Kub88a,Kub88b}.

The calorimeter consists of 240 pure CsI crystals.  The chosen
geodesic division results in 220 truncated hexagonal and pentagonal
pyramids covering the total solid angle of 0.77$\times$$4\pi$\/ sr. An
additional 20 veto crystals line two detector openings for beam entry
and detector readout, and act as electromagnetic shower leakage
vetoes.  The inner radius of the calorimeter is 26$\,$cm, and the
radial module length is 22$\,$cm, corresponding to 12 CsI radiation
lengths ($X_0$=1.85$\,$cm~\cite{PDG}). There are nine different module
shapes: four irregular hexagonal truncated pyramids (we label them
HEX--A, HEX--B, HEX--C, and HEX--D), one regular pentagon (PENT), and
two irregular half-hexagonal truncated pyramids (HEX--D1 and HEX--D2),
plus two trapezohedrons which serve as calorimeter vetoes (VET--1 and
VET--2).  The CsI module volumes vary from 797$\,$cm$^3$ (HEX--D1/2)
to 1718$\,$cm$^3$ (HEX--C).

All CsI crystal surfaces were polished and hand-painted with a special 
organo-silicon mixture in order to optimize the light collection and minimize
surface deterioration~\cite{Gor00}.

The average scintillation light yield of the CsI detectors measured in
a dedicated apparatus was 62 photoelectrons/MeV at the ambient
temperature of 18$^\circ$C. The average single CsI detector timing
uncertainty was determined to be 0.68$\,$ns rms. The temperature
coefficient of the light output extracted from cosmic muon
measurements was $-1.56\,\%/^\circ\,$C~\cite{Frl00,Frl01a}.

\bigskip
\section{Photomultiplier tubes, voltage dividers, and high voltage supplies}\label{sec:pmt}
\medskip

Electron Tubes (ET, formerly EMI) 9821QKB 10-stage fast
PMT's~\cite{EMI} with $\diameter$75$\,$mm end windows were glued to the
back faces of the hexagonal and pentagonal CsI crystals using a 300
$\mu$m layer of Sylgard 184 silicone elastomer (Dow Corning RTV
silicon rubber plus catalyst). The smaller half-hexagonal and
trapezial detector modules were equipped with $\diameter$46\,mm
10-stage ET/EMI 9211QKA PMT's~\cite{EMI}. Both types of PMT have
quartz windows transparent to light with wavelengths down to
175$\,$nm.  The window transparency peaks near
$\sim\,$380$\,$nm~\cite{EMI} and is therefore approximately matched to
the spectral excitation of the pure CsI fast scintillation light
component, which has maximum room temperature emission near
$\sim\,$310 nm~\cite{Woo90}.

The PMT high voltage dividers, designed and built at the University of
Virginia, were based on an ET-recommended circuit.  The dividers reduce
the so-called ``super-linearity'' exhibited by many PMT's well below
the onset of saturation ($>50\,\mu$A) by minimizing the ratio of anode
current to resistor string current with adequately sized capacitors on
the last few dynodes and through use of active circuit elements.  The
diagram of the voltage divider circuit is shown in
Fig.~\ref{fig:base}.

We have chosen the SbCs dynodes offered by ET to suppress
the dynode material-dependent PMT gain shift that occurs at small
anode currents (1--10$\,\mu$A).  The maximum PMT non-linearity
measured in a test with a pair of light-emitting diodes was less than
2$\,$\% over the full dynamic range expected in the $\pi_\beta$\/
decay rate measurement~\cite{Col95}.

Two LeCroy 1440 high voltage mainframes provided the high voltage for
all PMT's.  This 340-channel HV system had computer control capability
and thus allowed for frequent, remote changes in the HV supplied to
the PMT's.  The demand HV values could be set in 1\,V increments with
an accuracy and reproducibility of $\simeq$\,1\,V, which corresponded
to a gain change of $\sim$\,0.5\% for our CsI detector PMT's operating in
the range of 1500\,V to 2200\,V.

\bigskip
\section{Gain-matching procedure}\label{sec:match}
\medskip

\subsection{Calorimeter energy clustering and clumping}\label{sec:clust}
\medskip

Energies deposited by charged and neutral particles in the calorimeter
are used not only to measure the energy of the traversing particle but
also to define a basic element of the PIBETA detector trigger
logic. Therefore, the signal coming from the calorimeter PMT's was
split into two branches, namely a ``trigger'' branch which supplied
the signal for the trigger logic and the analog ``signal'' branch at
the input of the digitizing FASTBUS ADC's.  In order to obtain the
best possible performance in both branches we should consider the
following characteristics of the calorimeter response to 65--70\,MeV
photons from $\pi_\beta$ decay at rest and 69.8\,MeV positrons from
$\pi^+\to e^+\nu$ decay process:
\begin{enumerate}
\item[(1)] Electromagnetic shower profiles of the mean deposited
energies are similar for photons and positrons, in particular for
$\theta_c\ge 12^\circ$, with $\theta_c$ being the half-angle of a
conical bin concentric with the direction of an incident particle.
\item [(2)] A group of 9 detectors (a CsI ``cluster'') constitutes an
excellent summed energy trigger as it registers on average $\ge$\,98\%
of the incident particle energy.
\item[(3)] In case of a non-central hit, the sum of the energies in the
crystal with the most deposited energy and its nearest neighbors 
contains 90\,\% of the entire calorimeter energy.
\end{enumerate}

The summing and discrimination of the analog CsI ``cluster'' pulses in
the trigger branch was done in dedicated linear summer/discriminator
modules.  The logic signal outputs of these custom made single-width
NIM UVA125 summer/discriminator units provided the basic building
blocks for more complex physics triggers.

To analyze an acquired event we make use of signal-branch ADC and TDC
values recorded for ``clumps,'' CsI detector groupings consisting of
the centrally hit crystal and its nearest neighbors.  The initial
inspiration for our algorithm was the Crystal Ball detector idea of
``a clump discriminator function''~\cite{Ore80}.  The cluster-finding
algorithm operates in the on-line analyzer program and identifies
calorimeter shower clumps due to the interaction of a single ionizing
particle.  The algorithm first constructs a list of up to 6
nonadjacent calorimeter modules in order of decreasing deposited
energy (``clump centers'').  The minimum calibrated energy allowed for
a clump center is 1.0$\,$MeV, which is lower than the low energy
discriminator threshold (LT) of $\simeq\,$4.5\,MeV.  The clump
energies are next calculated by adding the energies of neighboring
modules to the energy recorded in the clump center, provided that, the
TDC hit values of the neighbors fall within a specified time window of
10$\,$ns.  Energies of crystals outside that time window are not
included in the energy sum, as they are assumed to be related to
accidental coincidences.  The number of nearest neighbors varies from
5 for a centrally hit pentagon, to 7 neighbors for the outlying HEX-C
modules.  The time associated with a clump is calculated as the
energy-weighted average of all clump members.  The clumping algorithm
finally saves summed energies, times and the relative angles between all
clump pairs and these clump sums are used in the gain matching process.

Hence, energy calibration of the {\sc PIBETA} calorimeter involves two
correlated processes: (1) equalizing the response of 220
trigger-defining non-veto CsI detector signals to a common UVA125
discriminator threshold by means of adjusting high voltages applied to
individual CsI PMT's, and (2) calibrating analog signal gains of 220
non-veto CsI detectors to achieve best possible energy resolution in
the signal branch, by introducing 220 software gain factors for
individual detectors.  The need for the software gains arises from the
relative difference between the signal amplitudes of an individual CsI
detector at the trigger branch and at the signal branch due to
different signal attenuation in the delay cables of two branches,
differences in the resistor values in passive signal splitters and/or
small mis-matches in time offsets of the two branches.  These two
adjustments can be done both manually by the experiment operators, as
well as automatically by a computer program.

\subsection{Threshold-matching CsI detectors}\label{sec:thr}
\medskip
As explained above, the first step in the gain equalization of the CsI
detectors requires the matching of energy-equivalent discriminator
thresholds of the individual modules.  For that purpose we have used
$\pi^+\to e^+\nu$ events collected with the one-arm high threshold
(HT) trigger set at $\simeq$54\,MeV.  We defined an on-line
one-dimensional 240-bin histogram for the indices of the individual
CsI detectors that were hit in an event. For every identified positron
track the histogram was filled with an index of the CsI clump center
(detector receiving the maximum energy deposition).  Ideally, for
perfectly threshold-matched detectors the resulting histogram shape
would reflect the solid angle covered by each detector.  Detector
HV's are therefore adjusted remotely via computer in an iterative
process until the threshold histogram shape matches the detector
yields predicted in a Monte Carlo simulation.

An example of four consecutive iterations in the threshold-matching
process is shown in Fig.~\ref{fig:thr}. At the end of the process the
HV's of the detectors were required to change by less than 1\,V on
average.  That operation fixed all CsI discriminator thresholds,
expressed in absolute MeV-equivalent units, at the same energy value
(usually within 0.5\,\%, corresponding to $\pm$1\,V in PMT HV).


\subsection{Gain-matching with $\pi^+\to e^+\nu$ peaks and Michel edge
  spectra}\label{sec:pi2e} 
\medskip

The software gain multipliers which optimize the energy resolution of
the digitized ADC spectra can be determined once the discriminator
thresholds are equalized at the trigger branch by adjusting the PMT
high voltages.  For a detector equipped with an $n$-stage PMT, a
normalized gain change $g$, relating two different gain settings 1 and
2, depends on the ratio of the software gains $s_i$ and the
corresponding high voltages ${\rm V}_i$:
\begin{equation}
g = {{s_1} \over {s_2}}\cdot
\left({{\rm V_1} \over {\rm V_2}}\right) ^m,
\label{eq:1}
\end{equation}
where $m$ is an exponent close to $n$, the number of the PMT
stages. Parameter $m$ depends on the design of the PMT used, as well
as on its voltage divider.  The above equation gives the real gain of
a scintillator detector at the setting 2, related to the gain at
setting 1.  The actual detected energies $E_i$ are then calculated
from the raw ADC$_i$ values:
\begin{equation}
E_i ({\rm MeV})=s_i\cdot{\rm ADC}_i.
\label{eq:2}
\end{equation}
Figure~\ref{fig:csi_hv} represents the gain measurements of a
representative CsI detector as a function of the high voltage applied
to its PMT.  The fitted coefficient $m$ is equal to 9, consistent with
the expectation for a 10-stage photomultiplier. The distribution of
the gain exponents for a large subset of the CsI detectors is shown in
Fig.~\ref{fig:csi_data}.  The average value of the parameter $m$ is
9.4 with the rms spread of 1.4 units.

 As indicated in the Introduction, we have used both (i) the
continuous $\mu^+\to e^+\nu\bar{\nu}$ (Michel) positron spectrum, and
(ii) the mono-energetic $\pi_{e2}$ peak in order to determine the HV
change and software gain change applied to each CsI module.  In
Fig.~\ref{fig:michpe2} we show the GEANT3~\cite{Bru94} Monte Carlo
simulation of the full Michel energy spectrum taken with the low
threshold trigger (top panel) and the $\pi_{e2}$ positron recorded
with the high threshold trigger (middle panel). The MC simulation
reflects a realistic and detailed description of the PIBETA detector
response.  Mixing both energy distributions with the $\pi^+\to e^+\nu$
branching ratio and simulating the realistic high threshold trigger
produces the bottom panel of Fig.~\ref{fig:michpe2}.  The high voltage
settings of all CsI detectors can be scaled simultaneously by a common
factor.  The effect of the overall HV re-scaling on our energy spectra
is demonstrated in three panels of Fig.~\ref{fig:fgainmatch}.  Similar
picture is observed for the energy spectrum in a clump centered around
any given non-veto crystal.  Therefore, such a combined positron
spectrum in the HT trigger can be used for the in situ gain-matching
of the entire CsI calorimeter. For the well-matched HT trigger the
Michel-edge-to-$\pi^+\to e^+\nu$ peaks ratio was empirically fixed to
be three-to-one. Keeping this ratio a constant provides continuous
monitoring of the trigger threshold, while the precise position of the
$\pi_{e2}$ decay positron energy peak determines the value of the
software gains to be applied to the individual crystals.

The MIDAS data acquisition system used in the experiment incorporates
an integrated slow control system with a fast on-line database (ODB)
and a history system~\cite{Rit01}.  Values of the software gains $s_i$
(all initially set to 1.0) are kept in the on-line experiment database
in the global memory section accessible to all running processes.  If
automatic gain-matching is enabled by setting the corresponding ODB
flag, counting statistics in the 220 energy clump histograms are
checked at the end of every run.  For the histograms with integrated
event counts exceeding the pre-set minimum yield of 1,000 events, the
$\pi_{e2}$ and Michel edge peaks are simultaneously fitted with a
double Gaussian function. The example of such a fit for four
representative CsI detectors (PENT 8, HEX-A 45, HEX-C 130, and HEX-D
190) is shown in Fig.~\ref{fig:4exam}.

Once at least 210 histograms have successful fits, the fit results are
used to calculate new values of software gains and to rewrite the old
values in the on-line database. The two-dimensional CsI energy
histogram shown in Fig.~\ref{fig:pi2e2d} was used during routine gain
monitoring.  The CsI detector index is displayed along the vertical
histogram axis, enabling a shift taker to notice easily the detectors
whose Michel edges and $\pi_{e2}$ peaks are out of alignment.  At the
end of each gain matching iteration, a log file documenting the
Gaussian fits was generated automatically and submitted to the
electronic logbook. The head portion of that log file looks like this:

{\scriptsize
\begin{verbatim}
0 N=1254 C2=3.377 CM=54.73 WM= 3.62 C=67.55 W=10.05 G=0.997 +- 0.001 R=0.242 HV=-1
1 N=1100 C2=2.781 CM=53.95 WM= 4.24 C=66.51 W= 9.69 G=1.007 +- 0.001 R=0.307 HV=+0
2 N=1178 C2=2.825 CM=54.94 WM= 4.70 C=66.37 W=10.94 G=1.009 +- 0.002 R=0.305 HV=+0
3 N=1261 C2=3.333 CM=54.22 WM= 4.95 C=66.61 W=11.75 G=1.006 +- 0.002 R=0.338 HV=+0
4 N=1149 C2=3.118 CM=53.91 WM= 4.23 C=66.23 W= 9.84 G=1.017 +- 0.002 R=0.267 HV=-1
5 N=1226 C2=2.763 CM=52.89 WM= 4.05 C=66.91 W= 8.90 G=1.001 +- 0.001 R=0.301 HV=+0
6 N=1174 C2=2.485 CM=54.25 WM= 5.30 C=67.55 W= 9.64 G=0.987 +- 0.001 R=0.375 HV=+1
...
\end{verbatim}}

The first column in the above list is the detector number, followed by
the integrated event count, and the six fitted parameters.  As
described above, the change of the software gain $g$ in the eighth
column is calculated from the deviation of the second fitted peak $C$
from the preset target value.  The last two items are $R$, a parameter
proportional to the $\chi^2$ of the fit, and the calculated PMT HV
change applied in order to keep the Michel-edge-to-$\pi_{e2}$ peaks
ratio fixed.  Only a few software gain iterations were necessary to
attain nearly optimal energy resolution of the calorimeter.  In
Fig.~\ref{fig:gainiterations} we show the one-arm calorimeter energy
spectrum at the start of the measurement and the improvement in the
resolution following the two passes of software gain changes.


\bigskip
\section{Temperature dependence of light output}\label{sec:temp}
\medskip

The photoelectron statistics in the CsI crystals and resulting gain
and energy resolution of the CsI detector are highly dependent on the
ambient temperature, making it imperative to maintain the temperature
as stable as possible.  Additionally, CsI is a hygroscopic compound
whose optical properties are degraded by absorbed moisture.  The
design goals for the temperature stabilization system were to
maintain:
\begin{enumerate}
\item[(1)] constant CsI calorimeter temperature of 22$^\circ\,$C,
stable to $\pm 0.02^\circ\,$C, ensuring gain stability of $\pm
0.02\,$MeV at 70\,MeV;
\item[(2)] relative humidity inside the detector thermal housing
$\le$\,50\%.
\end{enumerate}

The PIBETA calorimeter is enclosed in a thermal housing lined with
five 4\,cm thick Styrodur panels with a low heat conductivity.  A
recirculating chiller unit uses water as a cooling fluid and maintains
the temperature constant within $0.02^\circ\,$C.  During detector
operation two heater/fan units in the air pipes next to the cooling
heat exchanger are used in a feedback loop to regulate the air
temperature to within $0.02^\circ\,$C.  The air temperature is
measured just after both heaters, and the average CsI calorimeter
temperature is determined using eight sensors distributed uniformly
around the CsI sphere.

Figure~\ref{fig:temp} shows the variation of the CsI light output
caused by one instance of a failure of the detector temperature
control system.  The three panels show that Michel edges and
$\pi_{e2}$ peak positions scale down with rising temperature in the
absence of gain matching.  Alternatively, these results indicate that
the absolute energy threshold, expressed in MeV, is proportional to
the temperature.  The average CsI temperature coefficient extracted
from the 2004 data set (panel 4 of Fig.~\ref{fig:temp}) was $(-1.9\pm
0.3)\,\%/^\circ$C, consistent with our earlier (1997-1999)
measurements~\cite{Frl00,Frl01a}.

\bigskip
\section{Calorimeter energy resolution and gain stability}\label{sec:resol}
\medskip

Consistent applications of the gain stabilization procedures described
in the preceding sections provide very stable response of the
calorimeter over long periods of operation, as illustrated in
Fig.~\ref{fig:stab}.  The top panel shows the run-to-run values for
the fitted $\pi_{e2}$ peaks for one representative 2 month period.
The bottom panel displays the associated rms of the Gaussian fit for
the upper part of the $\pi_{e2}$ energy line-shape.  The peak position
was stable within $\pm$0.2\,MeV. While gain-matching was done manually
in the set-up stages of the experiment, the subsequent runs achieve
better energy resolution thanks to computer-controlled detector
operation and lower beam stopping rate.

The peak $\pi^+\to e^+\nu_e$ positron energy in CsI is affected by
energy losses in the active target, plastic veto scintillator, and in
the insensitive layers in front of the CsI crystals, positron
annihilation losses, photoelectron statistics of individual CsI
modules, and axial and transverse coefficients parameterizing the
nonuniformities of CsI light collection. Overall, the $\pi_{e2}$ peak
position was measured with absolute accuracy of $\pm 0.03$\,MeV.  The
best achieved rms fractional resolution $\Delta E/E$ was $4.5$\,\% at
67\,MeV. Contributions from the photoelectron statistics and the light
collection nonuniformities produce an rms of 3.6\,\%, while the gain
matching uncertainty adds an additional rms of 2.6\,\%.


\bigskip
\section{Results and conclusions}\label{res_con}
\medskip   
We have designed an in situ computer-controlled system for the gain
control of the PIBETA modular CsI electromagnetic calorimeter. We have
used the high-energy edge of the Michel spectrum as well as the
mono-energetic $\pi^+\to e^+\nu$ positron peak as the reference
points.  The computer program running at the end of each run adjusted
the PMT high voltages and/or the software gain factors in order to
compensate for the gain drifts and the slow continuous gain decrease
due to the CsI radiation damage and aging.  The system has
successfully matched the detector gains and controlled individual
detector HV settings in the long term, virtually free of human
intervention, maintaining the calorimeter rms energy resolution for
69.8\,MeV decay positrons at $\Delta E/E \simeq$\,4.5\%.

\bigskip
\section{Acknowledgments}
\medskip
This work is supported and made possible by grants from the US National
Science Foundation and the Paul Scherrer Institute.
 

\clearpage
 
\vspace*{\stretch{1}}
\begin{figure}[!tpb] 
\caption{Schematic cross section of the {\sc PIBETA} apparatus showing
the main components: forward beam counter (BC), two active collimators
(AC1, AC2), active degrader (AD), active target (AT), two MWPC's and
their support, plastic scintillator charged particle veto detectors
(PV) and PMT's, pure CsI calorimeter and PMT's.}
\label{fig:det}
\end{figure}           

\begin{figure}[!tpb] 
\caption{Panel (i): basic geometry of the pure CsI
shower calorimeter. The sphere is made up of 240 elements, truncated
hexagonal, pentagonal, and trapezoidal pyramids; it covers about 80\% of
4$\pi$ in solid angle.  Panel (ii): an individual supercluster and its
5-supercluster complement.  The {\sc PIBETA} calorimeter comprises 10
minimally-overlapping superclusters.}
\label{fig:ball}
\end{figure}

\begin{figure}[!tpb] 
\caption{The PMT high voltage dividers, designed and built at the
  University of Virginia, are based on a modified ET/EMI-recommended
  circuit.}
\label{fig:base}
\end{figure}

\begin{figure}[!tpb] 
\caption{Four consecutive iterations in the matching of CsI 
discriminator thresholds.  Individual CsI detector HV settings were 
adjusted until the count rates of the positrons with energies
exceeding $\sim\,$50\,MeV are roughly equalized. CsI veto detectors
220--239 were not included in the trigger.} 
\label{fig:thr}
\end{figure}

\begin{figure}[!tpb] 
\caption{Dependence of an individual CsI detector's gain factor on the
ET/EMI 9822QKB PMT high voltage. The LED signal level was adjusted to
produce 700\,pC at 2000\,V.}
\label{fig:csi_hv}
\end{figure}

\begin{figure}[!tpb] 
\caption{Distribution of slope coefficients for $\log$(HV) vs
PMT gain for CsI detectors.  The average fitted value of the coefficient
was 9.4 for 10-stage PMT's.}
\label{fig:csi_data}
\end{figure}

\begin{figure}[!tpb] 
\caption{Monte Carlo simulation of the calorimeter response to (i) the
continuous Michel positron energy spectrum with a theoretical
end-point of 52.8\,MeV (top), (ii) monoenergetic 69.8\,MeV $\pi^+\to
e^+\nu$ energy spectrum (middle), and (iii) combined simulated ADC
spectrum in the 1-arm high threshold trigger (bottom).  The GEANT3
calculation used a realistic response of the CsI calorimeter.}
\label{fig:michpe2}
\end{figure}

\begin{figure}[!tpb] 
\caption{GEANT3 detector simulation: effects of a gain shift on the
1-arm low-threshold energy spectrum (top plot) and the resulting gain
matched energy spectra for the 1-arm high-threshold trigger (bottom
plots, after applying software gains).}
\label{fig:fgainmatch}
\end{figure}

\begin{figure}[!tpb] 
\caption{Typical individual double Gaussian fits of the Michel edge and
$\pi_{e2}$ peak for four different calibrated energy clump histograms.}
\label{fig:4exam}
\end{figure}

\vspace*{\stretch{1}}
\begin{figure}[!tpb] 
\caption{Composite two-dimensional online calorimeter energy spectrum
for the one-arm high-threshold trigger.  The abscissa represents CsI
detector calibrated energy sum in MeV; the ordinate displays the clump
sum index number from 0-239.}
\label{fig:pi2e2d}
\end{figure}
\vspace*{\stretch{2}}
\clearpage

\vspace*{\stretch{1}}
\begin{figure}[!tpb] 
\caption{Two iterations of gain-matching: measured positron ADC
spectra acquired with the 1-arm high-threshold trigger.  The positron
energy represents a sum of pedestal-corrected ADC values for the
centrally hit CsI detector and its nearest neighbors. The initial
energy resolution of the $\pi\to e\nu$ peak is 8.0\,\% rms, the final
resolution is 4.9\,\% rms.}
\label{fig:gainiterations}
\end{figure}

\begin{figure}[!tpb] 
\caption{Dependence of the CsI detector gain factors on ambient
temperature in the range $23.0\pm 1.2$ degrees Celsius.  The PIBETA
detector is operated inside a temperature-controlled thermal
enclosure.}
\label{fig:temp}
\end{figure}

\begin{figure}[!tpb] 
\caption{Long-term stability of the CsI online detector gains.  The
top panel shows the $\pi^+\to\e^+\nu$ peak position for $\sim\,$2,000
runs covering a two month period.  The bottom panel represents the
corresponding online peak resolution.}
\label{fig:stab}
\end{figure}
\vspace*{\stretch{2}}
\clearpage
%
%


\vspace*{\stretch{1}}
\centerline{\psfig{figure=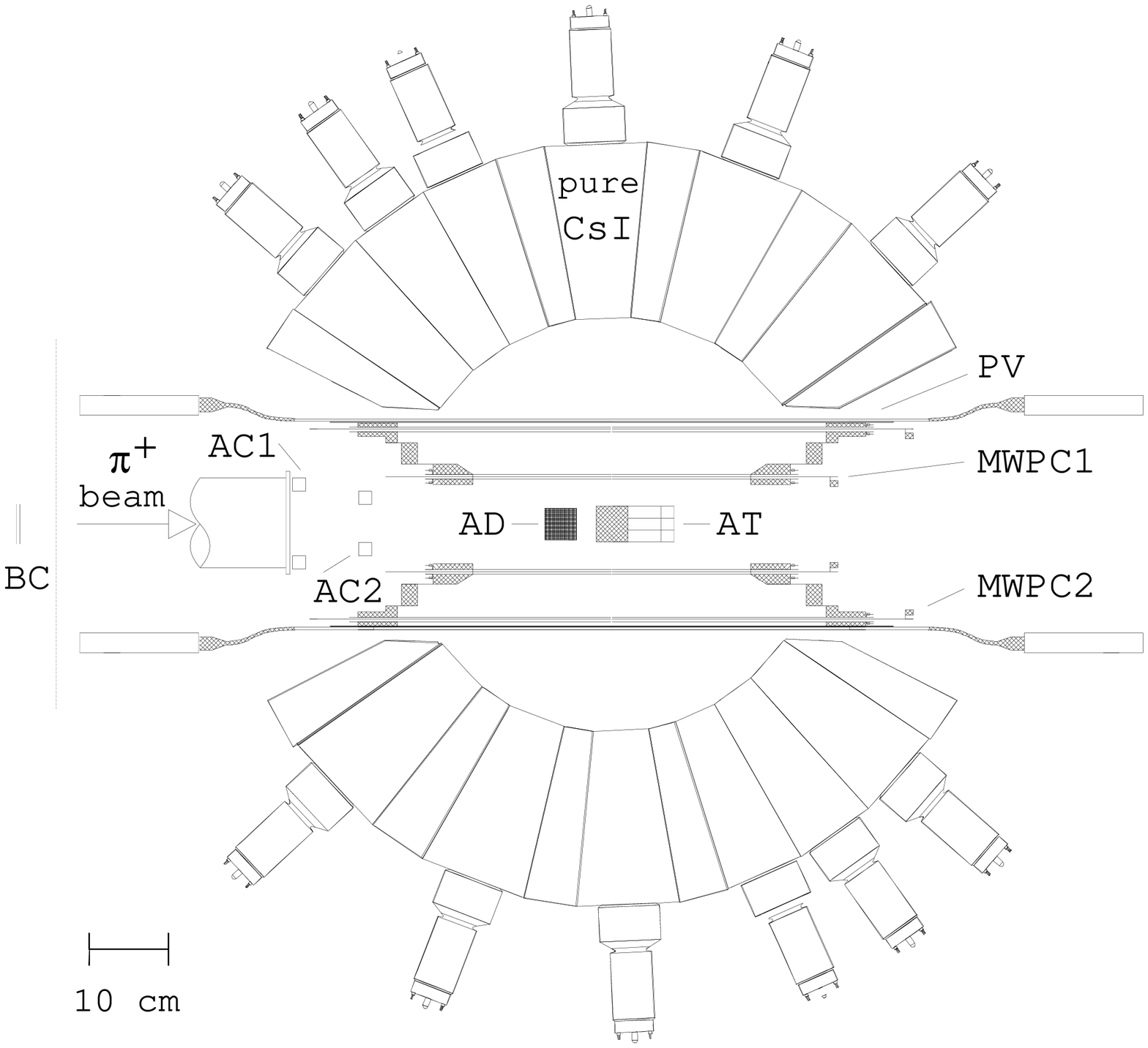,width=15cm}}
\vglue 1.5cm
\centerline{FIGURE 1}
\vspace*{\stretch{2}}
\clearpage    

\vspace*{\stretch{1}}
\hbox{\ }\noindent \hglue 3cm (i)
\centerline{\psfig{figure=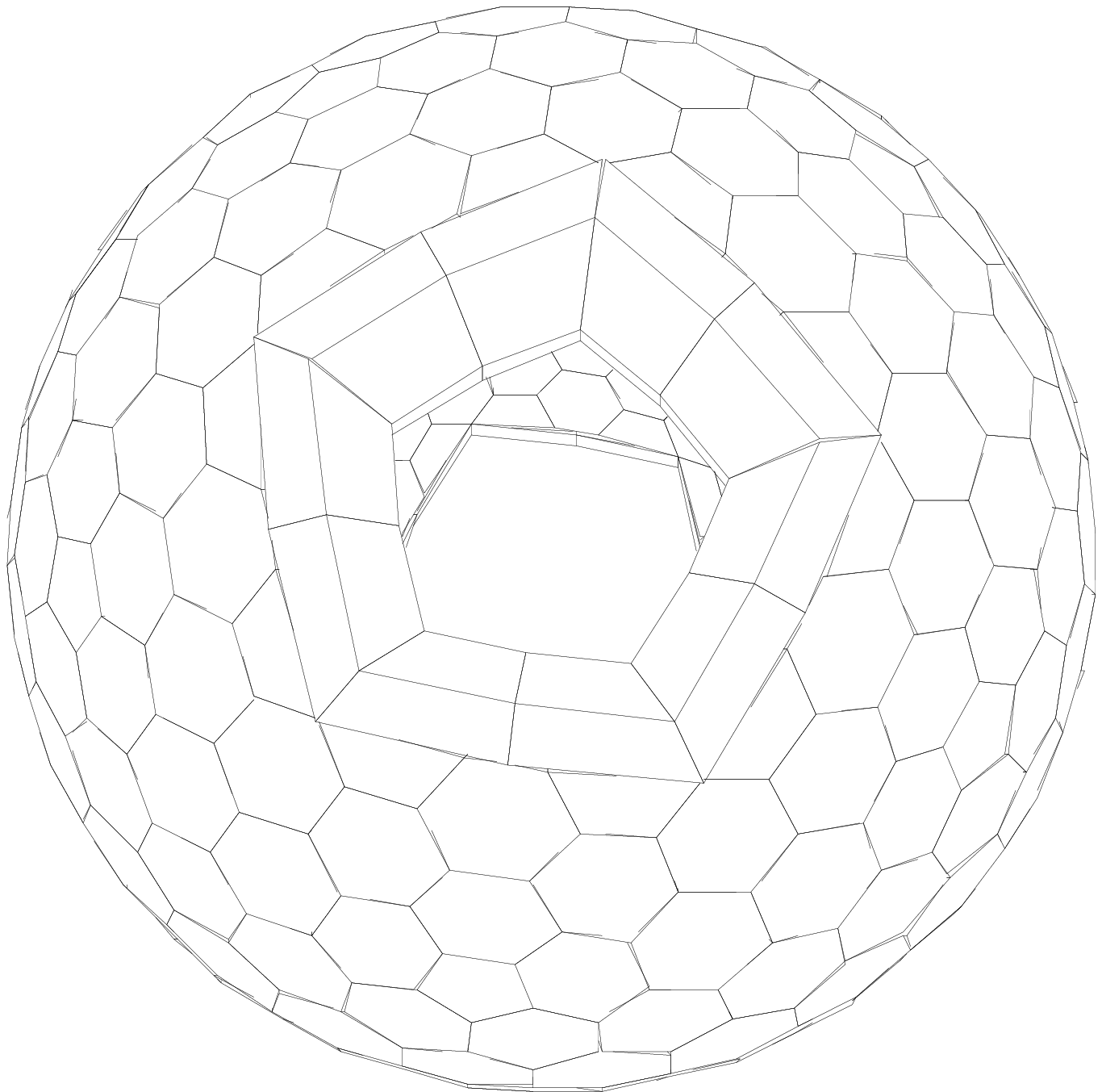,height=8cm}\hglue 3.5cm}
\vglue 1cm
\noindent \hglue 3cm (ii)
\centerline{\psfig{figure=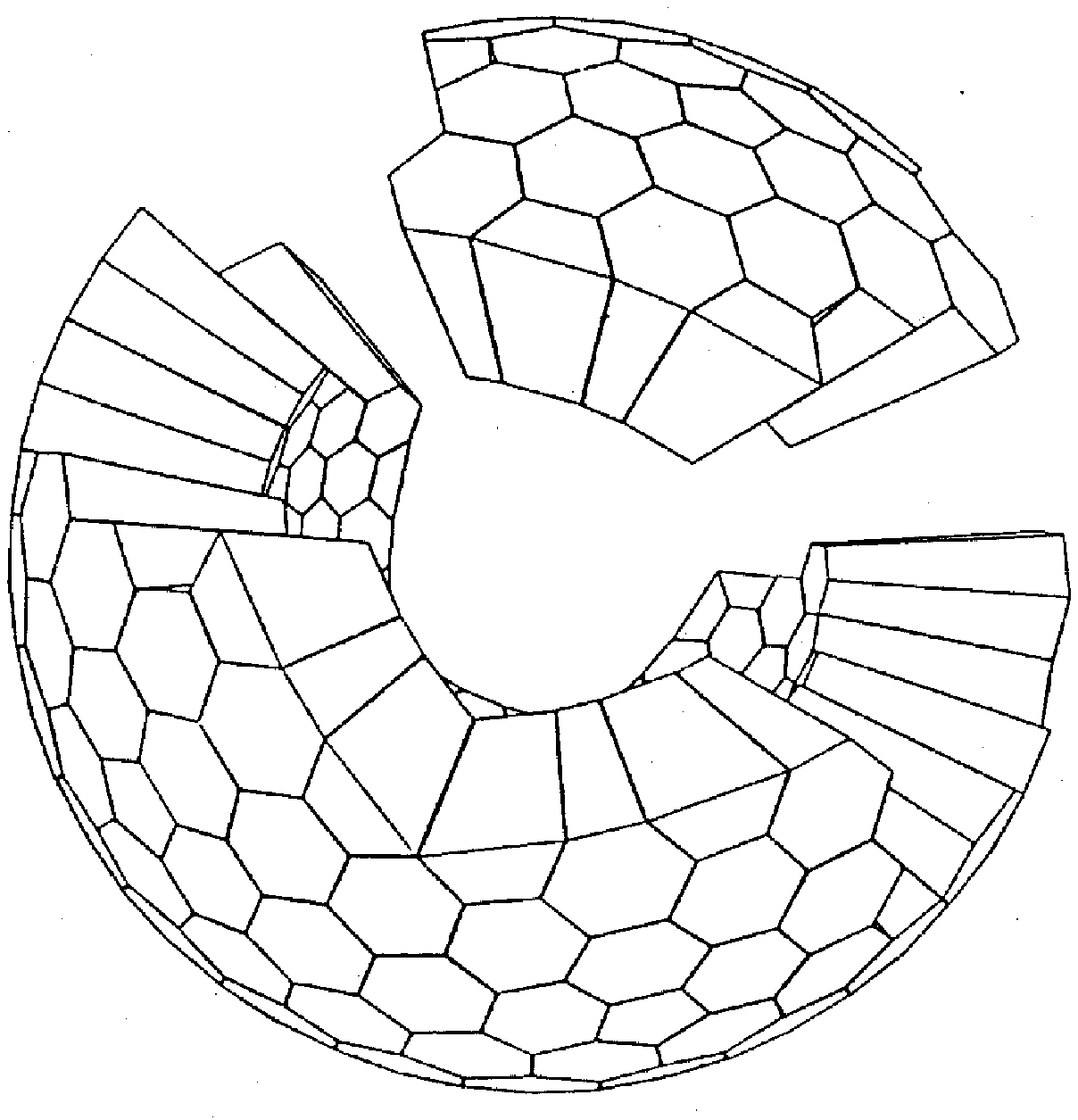,width=8.9cm}\hglue 3.5cm}
\vglue 1.0cm
\centerline{FIGURE 2}
\vspace*{\stretch{2}}
\clearpage    

\vspace*{\stretch{1}}
\centerline{\psfig{figure=fig3.eps,width=15cm}}
\vglue -3.5cm
\centerline{FIGURE 3}
\vspace*{\stretch{2}}
\clearpage

\vspace*{\stretch{1}}
\centerline{\psfig{figure=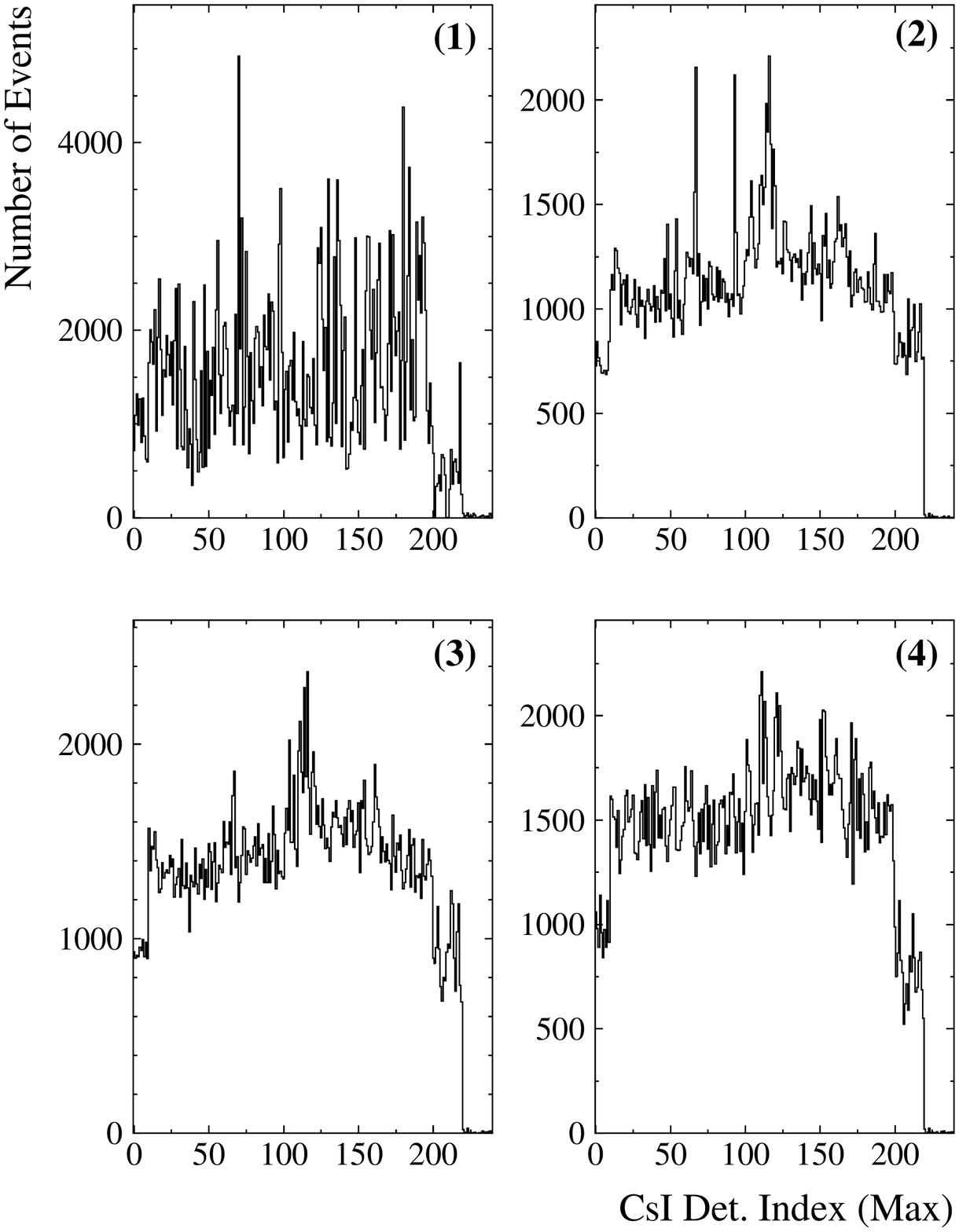,width=14cm}}
\bigskip
\centerline{FIGURE 4}
\vspace*{\stretch{2}}
\clearpage

\vspace*{\stretch{1}}
\centerline{\psfig{figure=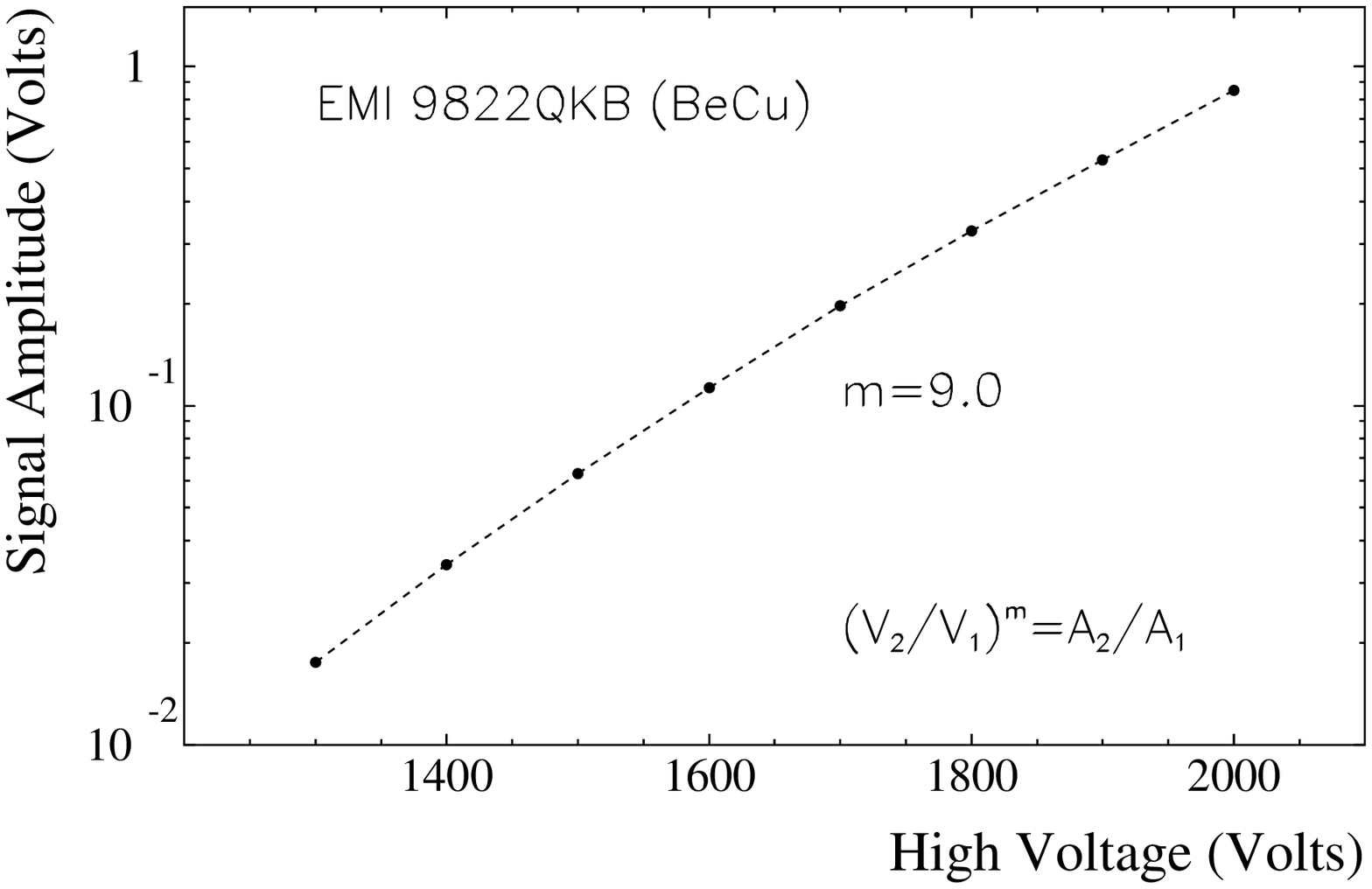,width=15cm}}
\vglue -8cm
\centerline{FIGURE 5}
\vspace*{\stretch{2}}
\clearpage

\vspace*{\stretch{1}}
\centerline{\psfig{figure=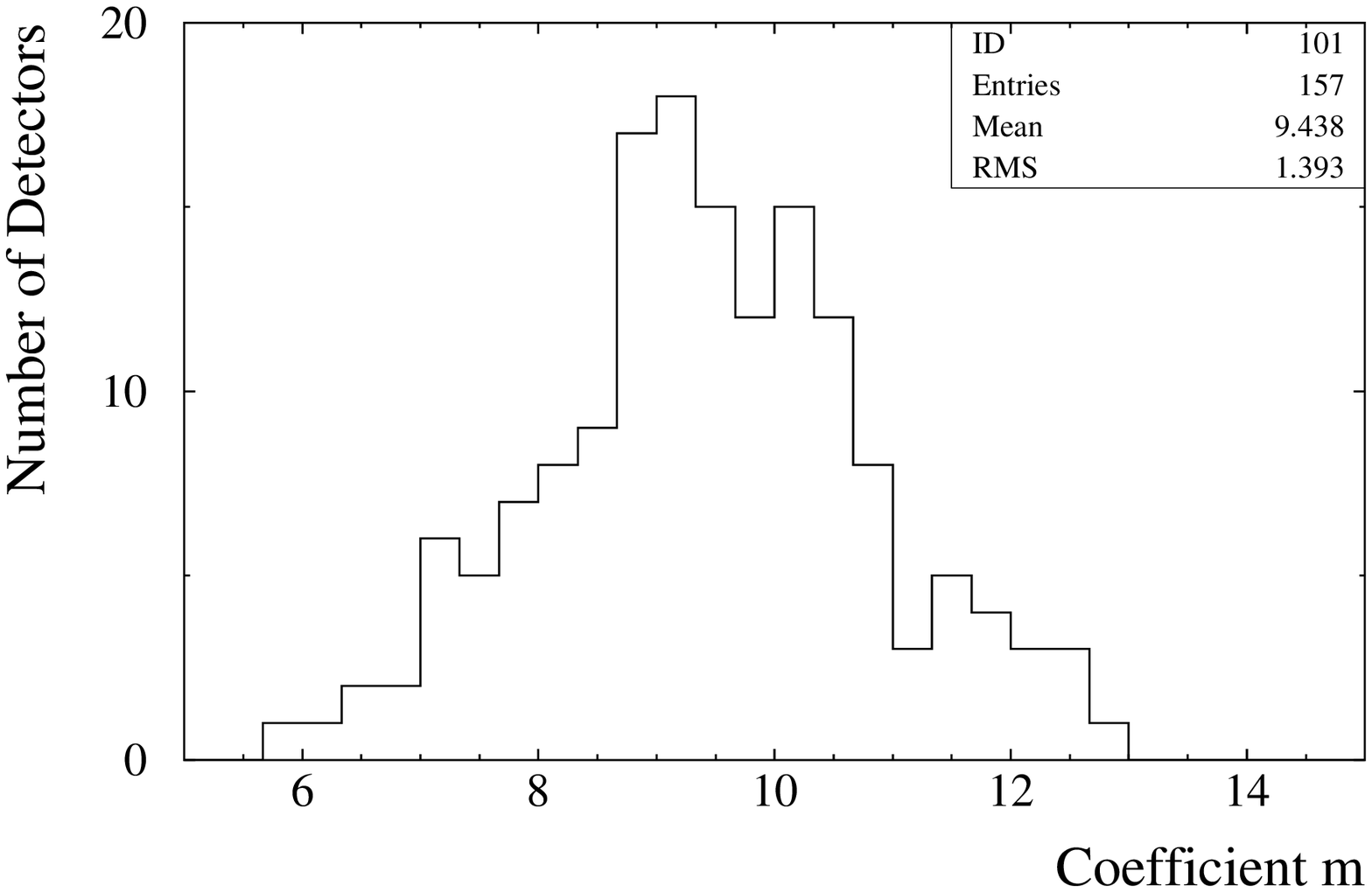,width=15cm}}
\vglue -8cm
\centerline{FIGURE 6}
\vspace*{\stretch{2}}
\clearpage

\vspace*{\stretch{1}}
\vskip -1cm
\centerline{\psfig{figure=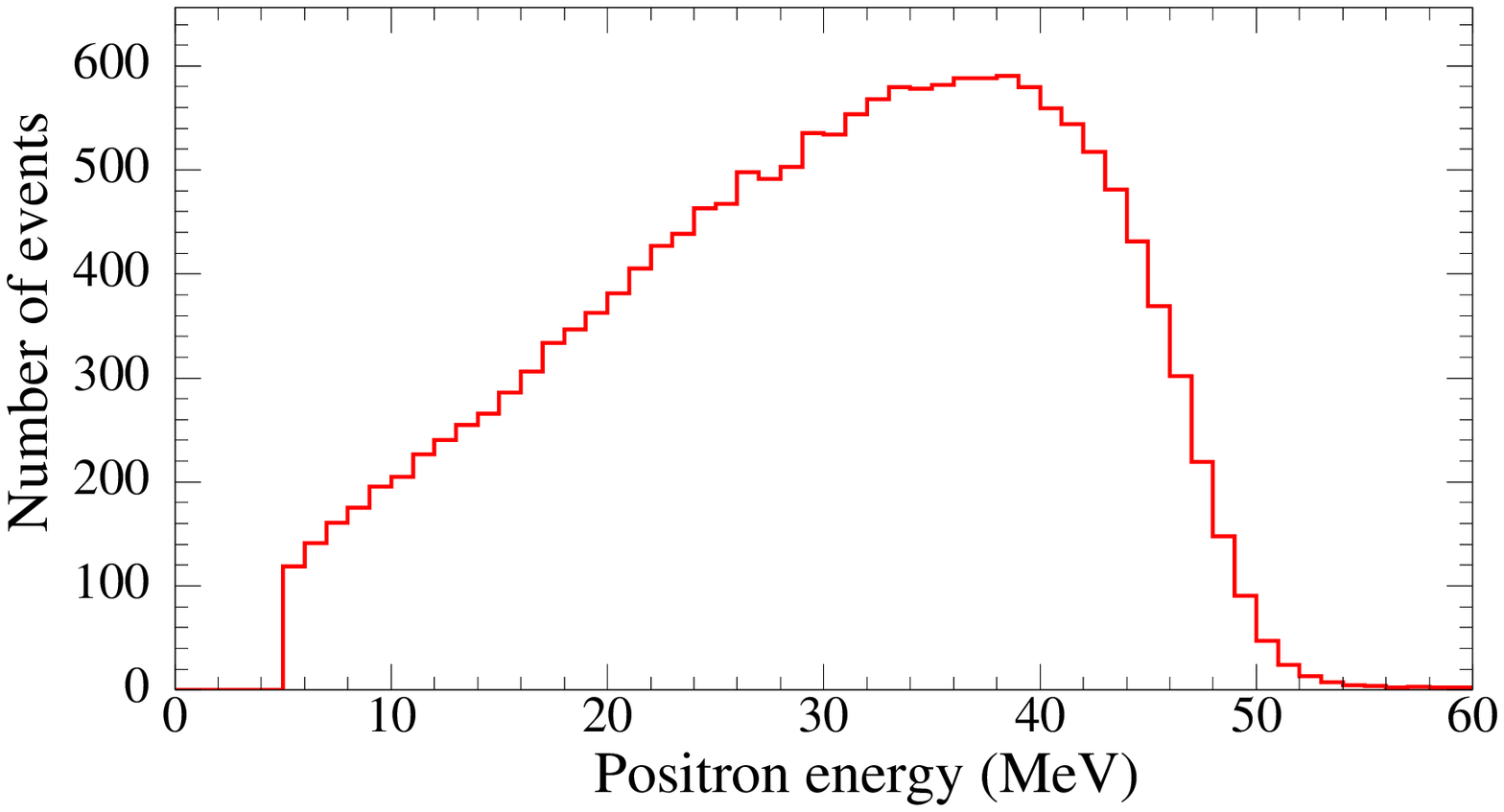,width=14cm}}
\centerline{\psfig{figure=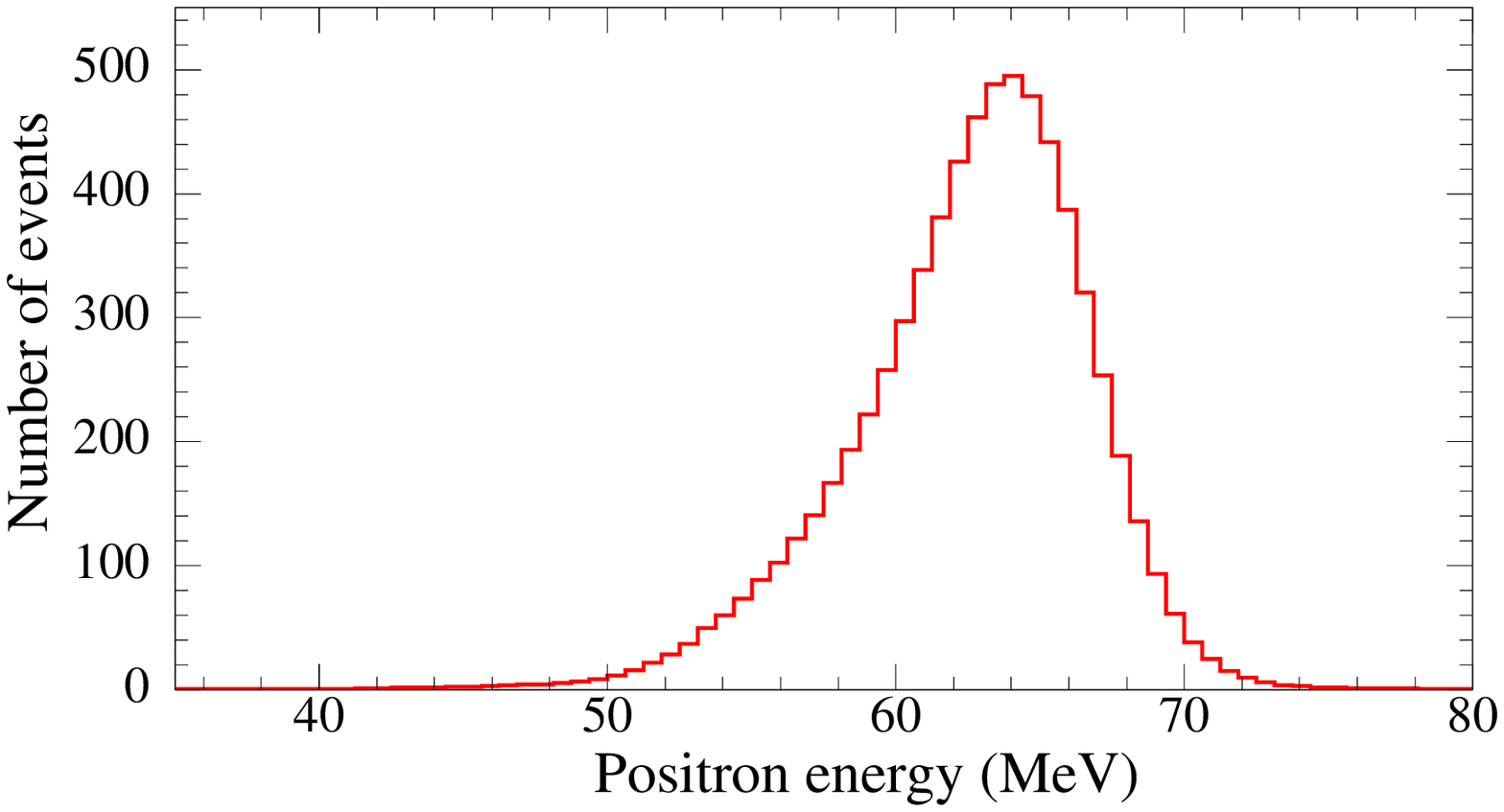,width=14cm}}
\centerline{\psfig{figure=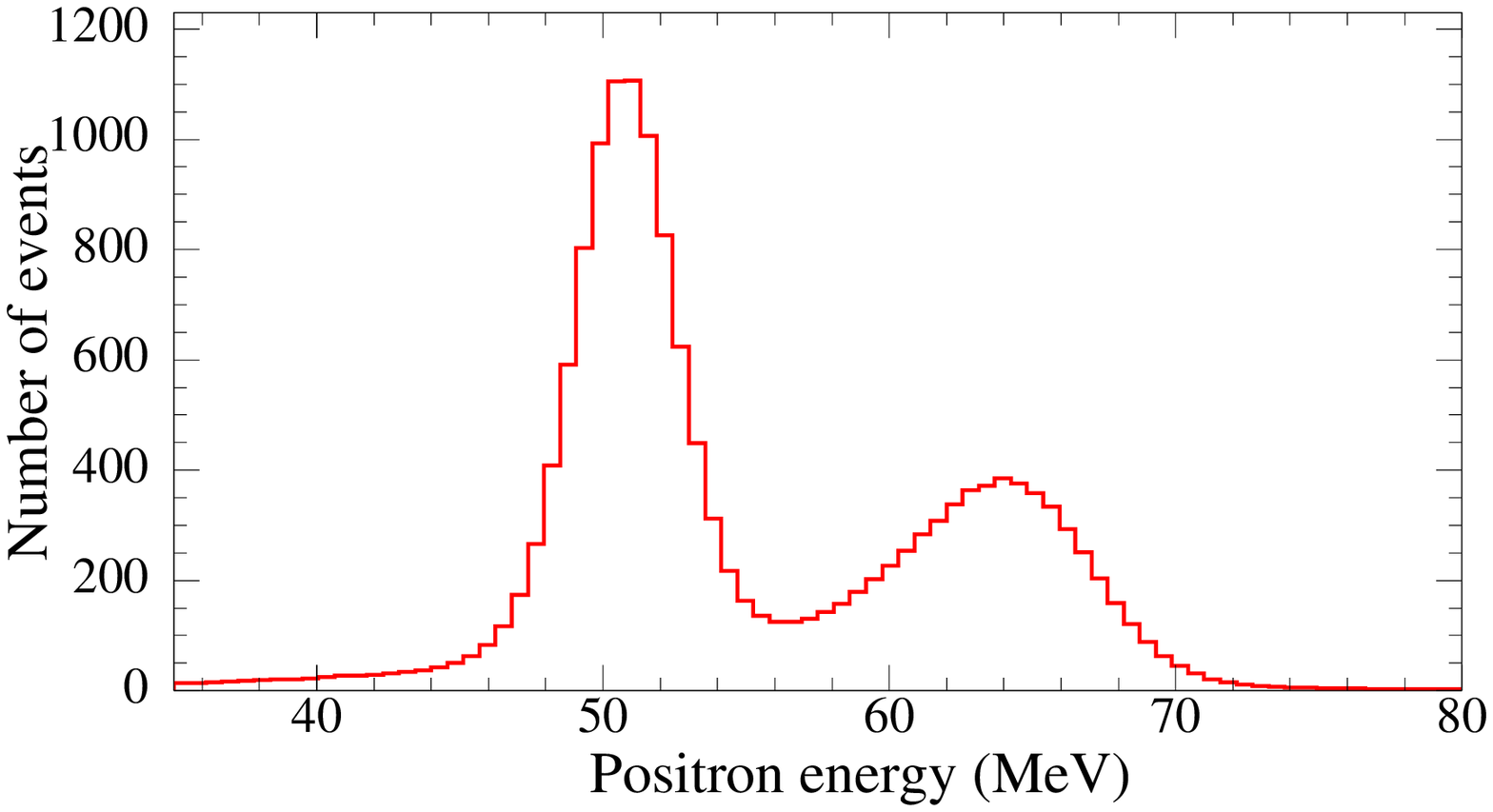,width=14cm}}
\vglue 0.5cm
\centerline{FIGURE 7}
\vspace*{\stretch{2}}
\clearpage

\vspace*{\stretch{1}}
\centerline{\psfig{figure=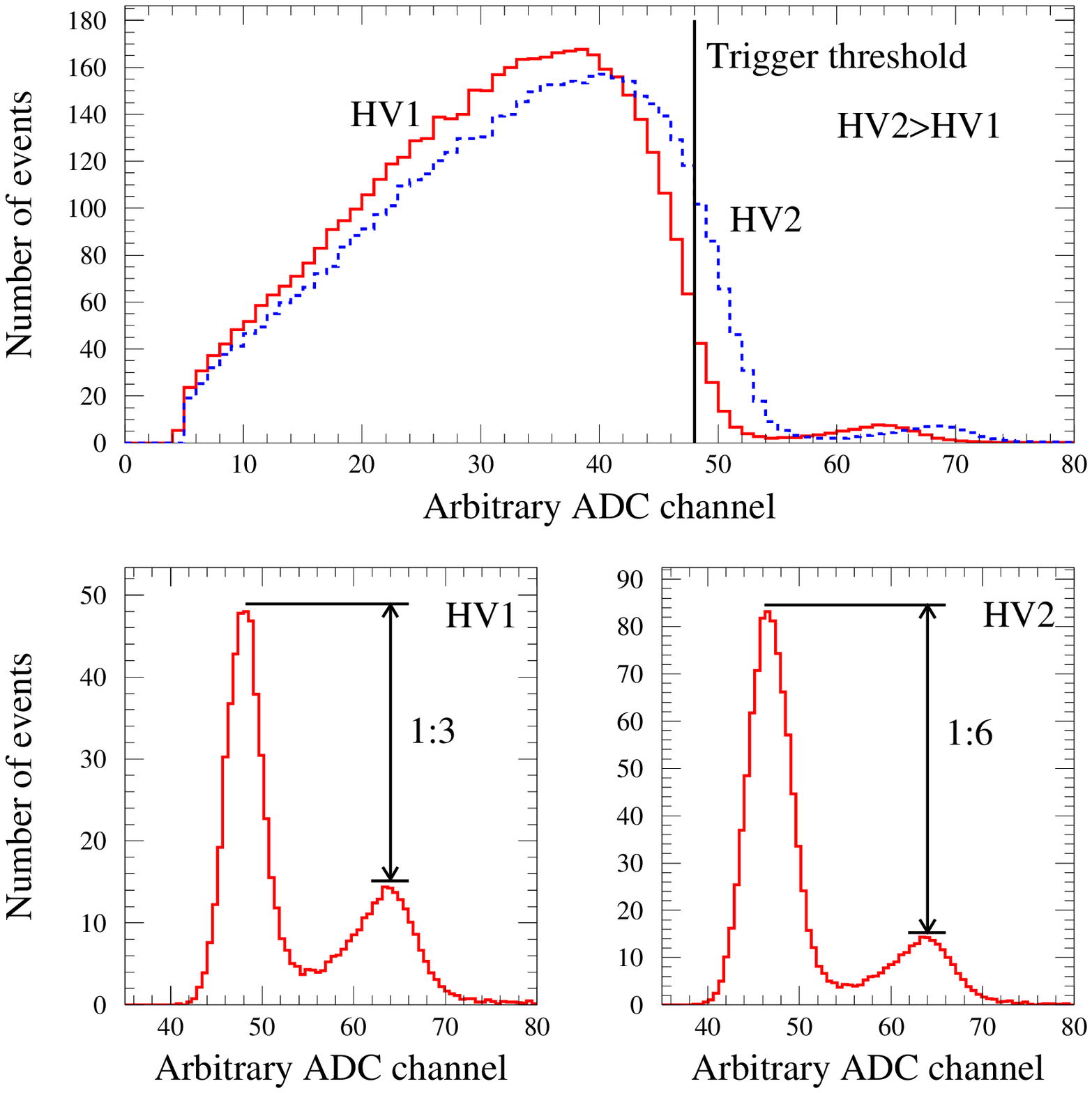,width=16cm}}
\vglue 1cm
\centerline{FIGURE 8}
\vspace*{\stretch{2}}
\clearpage

\vspace*{\stretch{1}}
\centerline{\psfig{figure=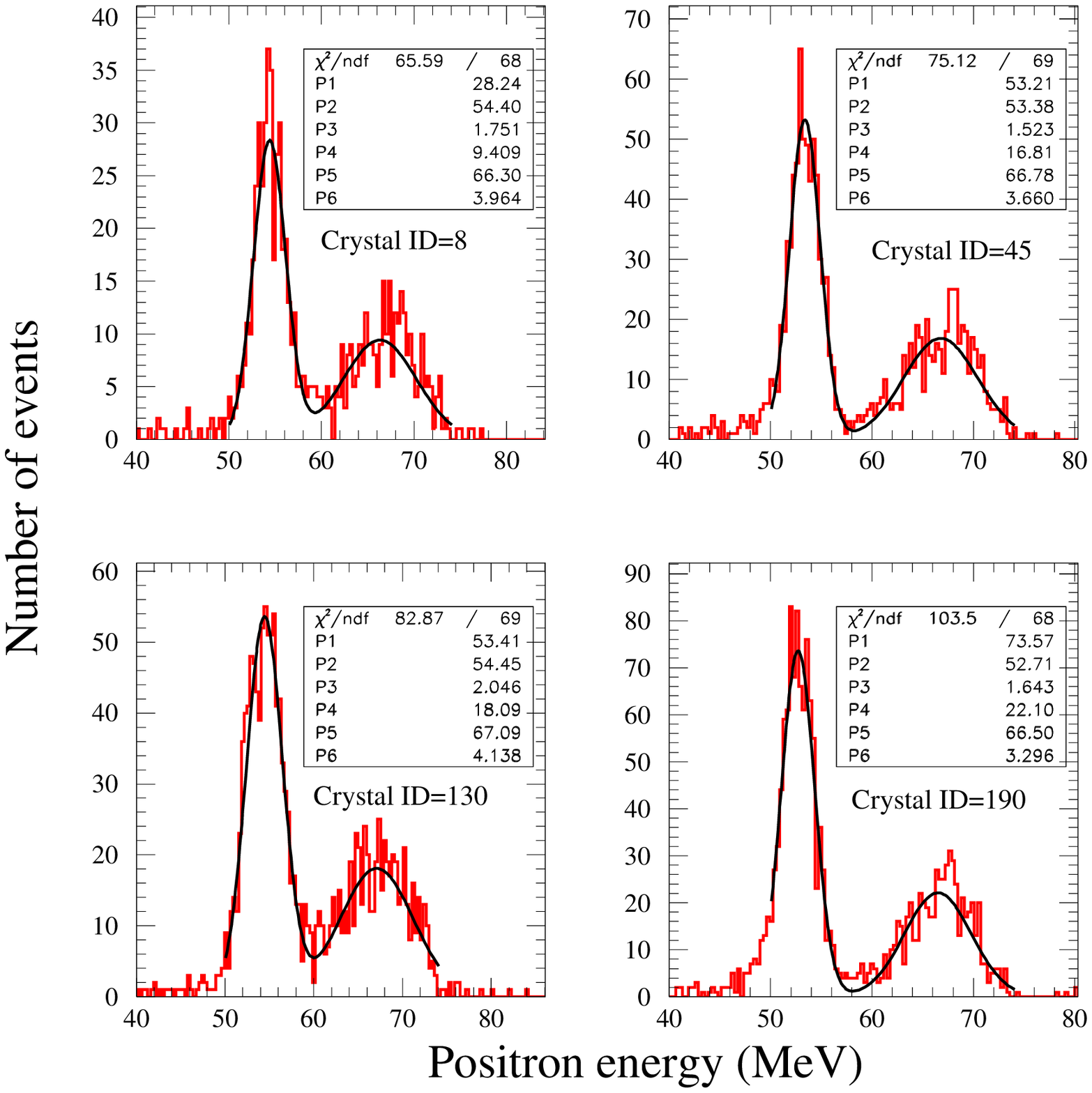,width=16cm}}
\vglue 1.5cm
\centerline{FIGURE 9}
\vspace*{\stretch{2}}
\clearpage

\vspace*{\stretch{1}}
\centerline{\psfig{figure=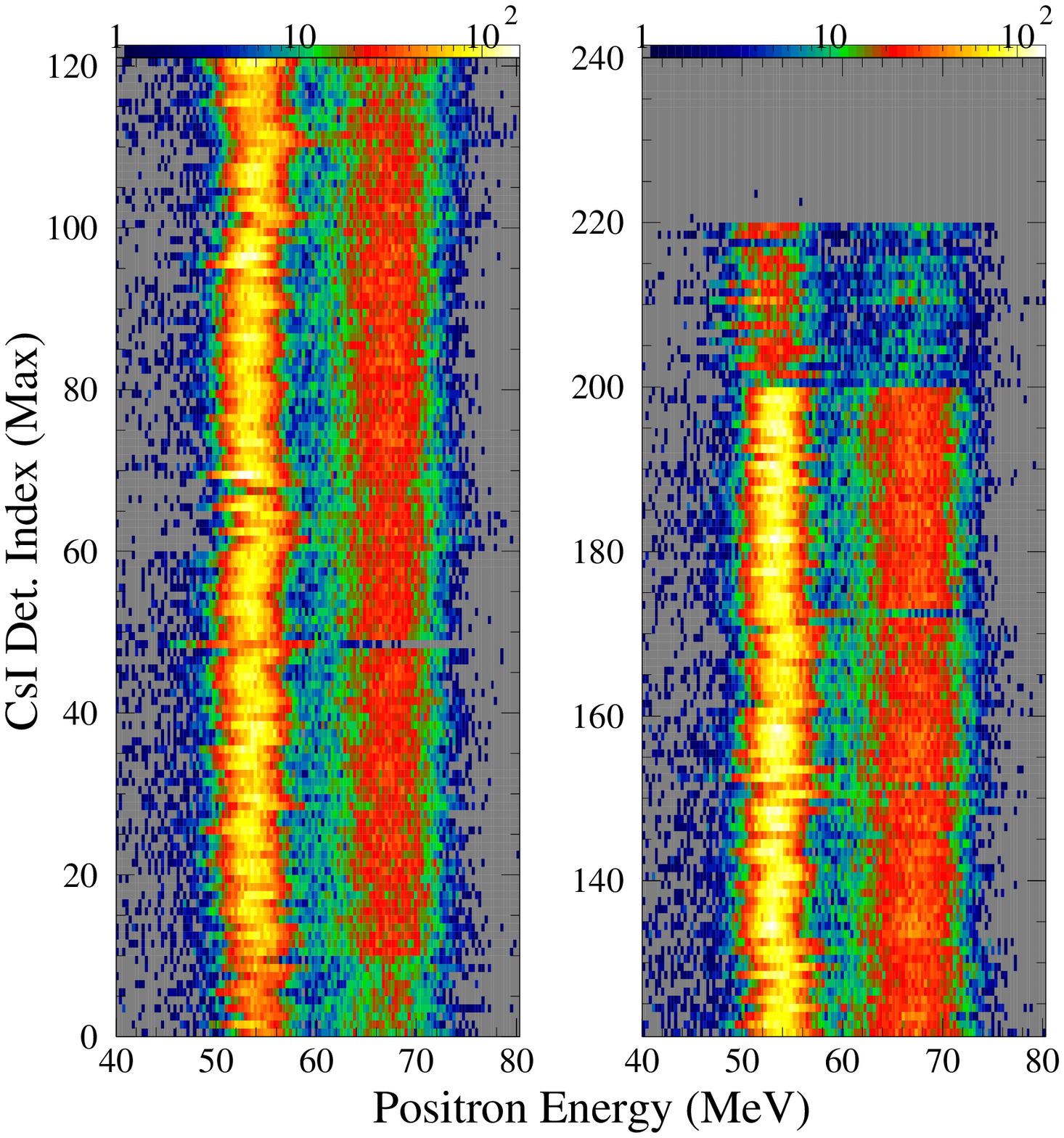,width=16cm}}
\vglue 1.5cm
\centerline{FIGURE 10}
\vspace*{\stretch{2}}
\clearpage

\vspace*{\stretch{1}}
\centerline{\psfig{figure=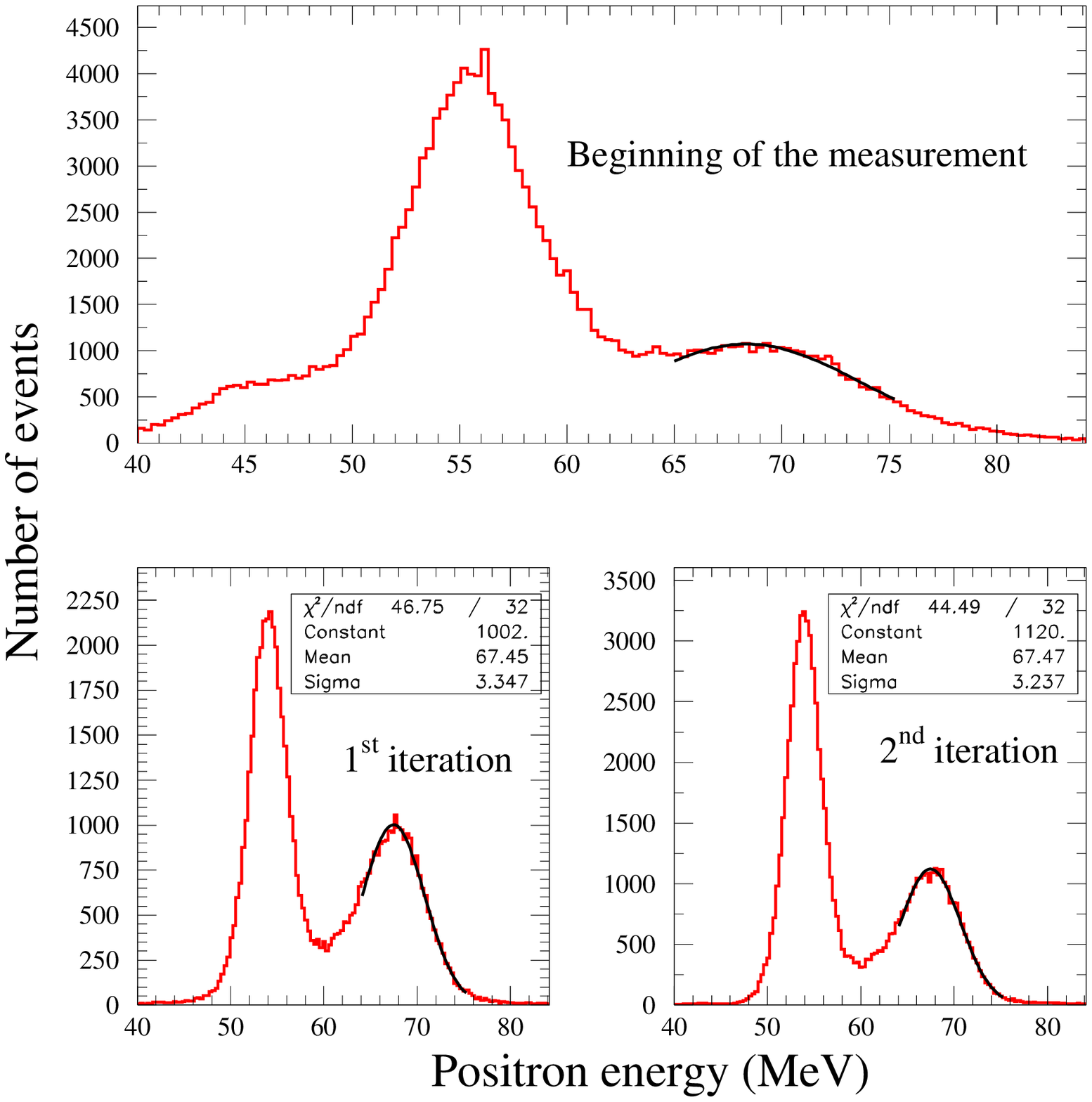,width=16cm}}
\vglue 1.5cm
\centerline{FIGURE 11}
\vspace*{\stretch{2}}
\clearpage

\vspace*{\stretch{1}}
\centerline{\psfig{figure=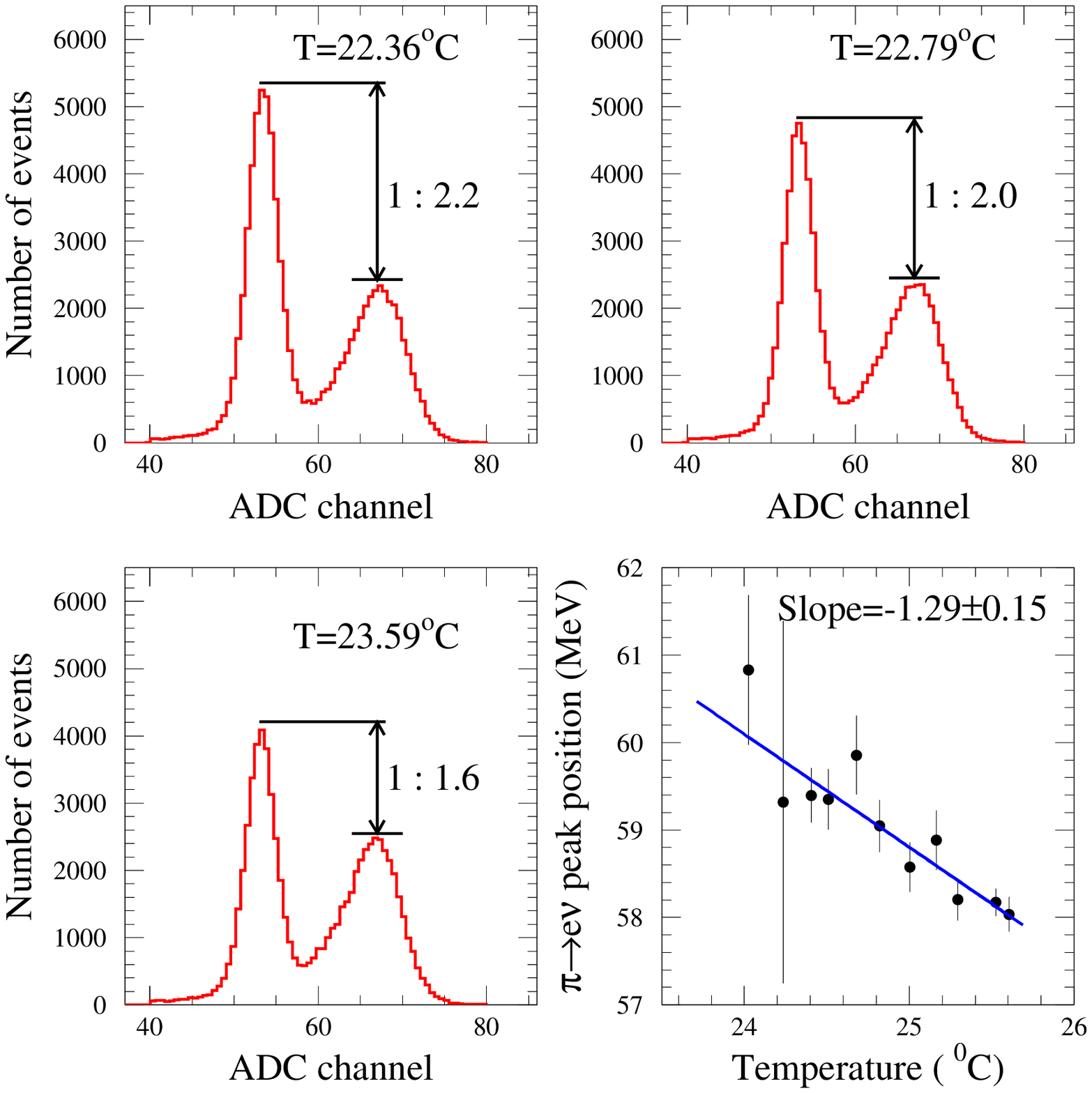,width=16cm}}
\bigskip\bigskip
\centerline{FIGURE 12}
\vspace*{\stretch{2}}
\clearpage

\vspace*{\stretch{1}}
\centerline{\psfig{figure=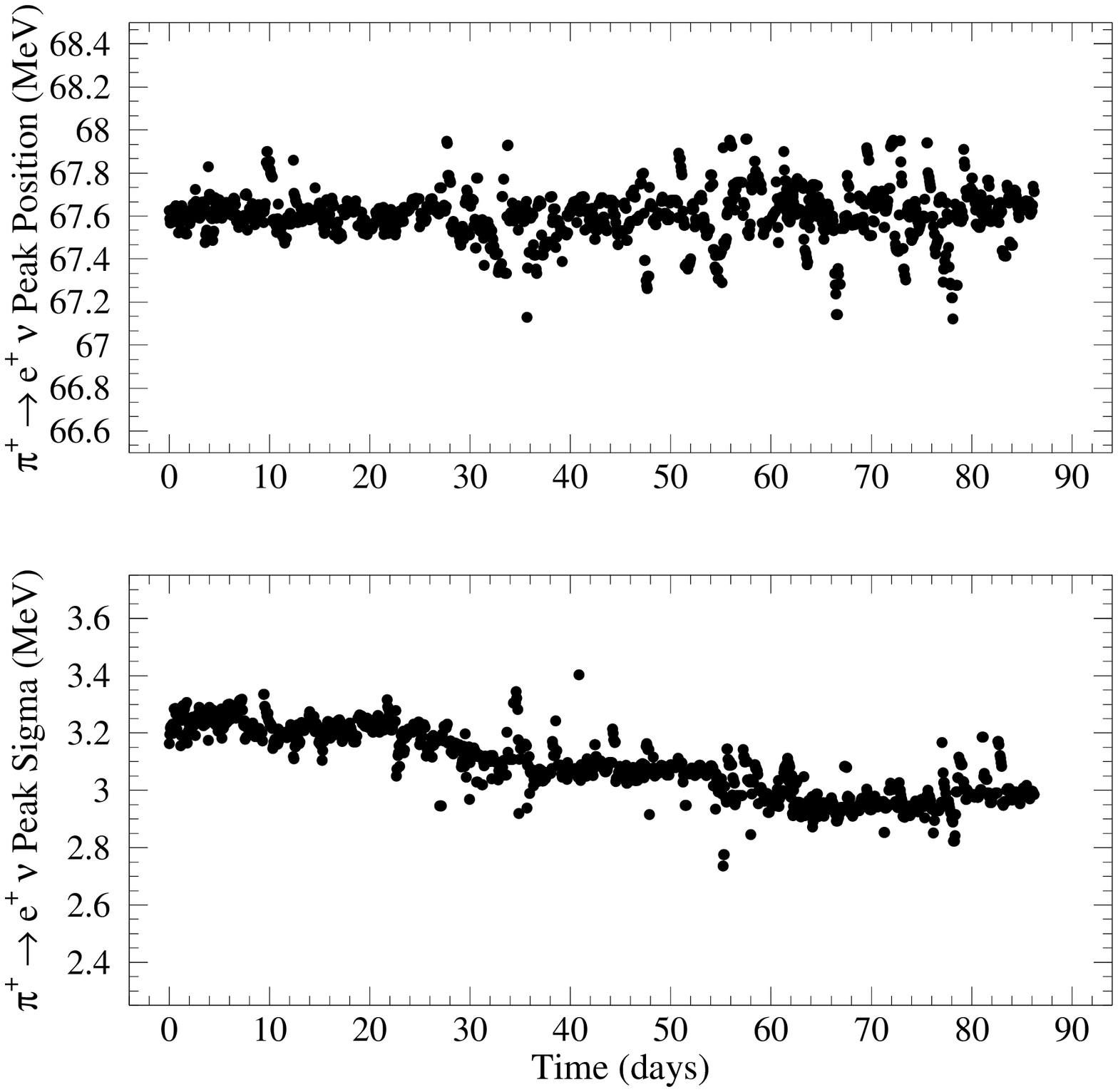,width=15cm}}
\bigskip
\centerline{FIGURE 13}
\vspace*{\stretch{2}}
\clearpage


\end{document}